# Pressure-induced Superconductivity in Zintl Topological Insulator SrIn$_2$As$_2$

Weizheng Cao[1#], Haifeng Yang[1,2#], Yongkai Li[3,4,5#], Cuiying Pei[1], Qi Wang[1,2], Yi Zhao[1], Changhua Li[1], Mingxin Zhang[1], Shihao Zhu[1], Juefei Wu[1], Lili Zhang[6], Zhiwei Wang[3,4,5*], Yugui Yao[3,4], Zhongkai Liu[1,2*], Yulin Chen[1,2,7], and Yanpeng Qi[1,2,8*]

1. School of Physical Science and Technology, ShanghaiTech University, Shanghai 201210, China
2. ShanghaiTech Laboratory for Topological Physics, ShanghaiTech University, Shanghai 201210, China
3. Centre for Quantum Physics, Key Laboratory of Advanced Optoelectronic Quantum Architecture and Measurement (MOE), School of Physics, Beijing Institute of Technology, Beijing 100081, China
4. Beijing Key Lab of Nanophotonics and Ultrafine Optoelectronic Systems, Beijing Institute of Technology, Beijing 100081, China
5. Material Science Center, Yangtze Delta Region Academy of Beijing Institute of Technology, Jiaxing, 314011, China
6. Shanghai Synchrotron Radiation Facility, Shanghai Advanced Research Institute, Chinese Academy of Sciences, Shanghai 201203, China
7. Department of Physics, Clarendon Laboratory, University of Oxford, Parks Road, Oxford OX1 3PU, UK
8. Shanghai Key Laboratory of High-resolution Electron Microscopy, ShanghaiTech University, Shanghai 201210, China

\# These authors contributed to this work equally.

\* Correspondence should be addressed to Y.Q. (qiyp@shanghaitech.edu.cn) Z.L (liuzhk@shanghaitech.edu.cn) or Z.W. (zhiweiwang@bit.edu.cn)

**ABSTRACT**

The Zintl compound $A$In$_2$$X$$_2$ ($A$ = Ca, Sr, and $X$ = P, As), as a theoretically predicted new non-magnetic topological insulator, requires experiments to understand their electronic structure and topological characteristics. In this paper, we systematically investigate the crystal structures and electronic properties of the Zintl compound SrIn$_2$As$_2$ under both ambient and high-pressure conditions. Based on systematic angle-resolved photoemission spectroscopy (ARPES) measurements, we observed the topological surface states on its (001) surface as predicted by calculations, indicating that SrIn$_2$As$_2$ is a strong topological insulator. Interestingly, application of pressure effectively tuned the crystal structure and electronic properties of SrIn$_2$As$_2$. Superconductivity is observed in SrIn$_2$As$_2$ for pressure where the temperature dependence of the resistivity changes from a semiconducting-like behavior to that of a metal. The observation of nontrivial topological states and pressure-induced superconductivity in SrIn$_2$As$_2$ provides crucial insights into the relationship between topology and superconductivity, as well as stimulates further studies of superconductivity in topological materials.

## INTRODUCTION

Topological insulator ( TI ) is a novel class of materials that possess both gapped bulk and exotic metallic surface states. The robust topological surface state ( TSS ) in TI has been confirmed by angle resolved photoemission spectroscopy ( ARPES ) [1, 2] and scanning tunneling microscopy ( STM ) [3-5], which also characterized by magneto transport such as Shubnikov – de Hass ( SdH ) oscillation [6, 7], etc. Due to the dissipationless flow of electric current at the surface, TI has been projected as promising candidates for developing next-generation, low-power-consuming, high-speed electronic, optoelectronic, and spintronic devices. The interplay between symmetry and topology in crystalline solids leads to other emerging topological quantum materials ( TQMs ), including topological crystalline insulators [8, 9], higher-order topological insulators [10, 11], Dirac [12, 13], Weyl [2, 14-16], nodal line [17, 18], and multifold semimetals [19, 20].

Since first discovery in the early 1900s by Eduard Zintl [21], Zintl compounds have been extensively studied for their fascinating physical properties, including superconductivity, colossal magnetoresistance, magnetic order, mixed-valence, thermoelectricity, and so on [22-28]. Several Zintl compounds have recently garnered attention due to their interesting topological properties. For instance, $EuIn_2As_2$ exhibits A - type antiferromagnetic ( AFM ) ordering at $T_N \sim 16$ K [29], and is predicted to be a promising candidate of the axion topological insulator [30]. ARPES results indicated a hole - type Fermi pocket around the Brillouin zone and the AFM transition accompanied by the axion insulator phase [31, 32]. The magnetotransport characterization revealed an anomalous Hall effect ( AHE ) originated from a nonvanishing Berry curvature and a large topological Hall effect ( THE ), which is attributed to the scalar spin chirality of the noncoplanar spin structure caused by the external field [33]. Another Zintl compound, $SrIn_2As_2$, which is isostructural with $EuIn_2As_2$, has been predicted to be a dual topological insulator with nontrivial $Z_2 = 1$ and mirror Chern number $C_M = -1$ via first principles calculations together with symmetry analysis [34]. Furthermore, $AIn_2X_2$ ($A$ = Ca, Sr, and $X$ = P, As) family exhibits a narrow band gap which can be easily tuned by high pressure or chemical doping [35-38]. Theoretical calculations indicated that the pressure-induced band gap reduces to zero at 6.6 GPa and then reopens, and $SrIn_2As_2$ exhibit a gapless non-trivial topological surface state, indicating a strong topological insulator [39]. However, these theoretical predictions have not been experimentally verified yet, and further investigation of the evolution of the physical properties of $AIn_2X_2$ under pressure is required.

In this work, we choose Zintl compound $SrIn_2As_2$, one member of the $AIn_2X_2$ family, to systematically investigate the crystal structures and electronic properties under both ambient and high-pressure conditions. The transport measurements reveal metallic behavior for $SrIn_2As_2$ with hole-dominated carriers. ARPES results demonstrate topological Dirac surface states on the (001) surface, which is consistent with previous theoretical calculation. Interestingly, superconductivity was observed at around 30 GPa and persisted approximately 2.1 K until the maximum experimental pressure. The pressure-induced structure phase transition in the Zintl compound $SrIn_2As_2$ is also discussed.

**EXPERIMENTAL DETAILS**

The single crystals of SrIn$_2$As$_2$ were grown by self-flux method. High-purity starting materials of Sr, In, and As were loaded in a quartz tube with the ratio of Sr : In : As = 1 : 12 : 2. The tube was sealed after it was evacuated to a vacuum of $2\times10^{-4}$ Pa. The raw materials were reacted and homogenized at 1373 K for several hours, followed by cooling down to 873 K at a rate of 2 K/h. Finally, Shiny crystal with apparent hexagonal edge was obtained after the indium flux was removed by centrifuge. The crystalline phase of SrIn$_2$As$_2$ was checked by the single-crystalline x-ray diffraction ( XRD, Cu $K_\alpha$, $\lambda$ = 1.54184 Å ). The chemical composition value of SrIn$_2$As$_2$ is given by energy-dispersive x-ray spectra ( EDX ). Transmission electron microscopy ( TEM ) measurements are carried out on a SrIn$_2$As$_2$ sample after high-pressure treated at 61.5 GPa. Electrical transport properties including resistivity, magnetoresistance, and Hall effect were performed on a physical property measurement system ( PPMS ).

In order to investigate the band structures of SrIn$_2$As$_2$, we performed systematic synchrotron-based ARPES measurements at BL03U of Shanghai Synchrotron Radiation Facility. Data were recorded by a Scienta Omicron DA30L analyzer. The measurement sample temperature and pressure were 15 K and $7\times10^{-11}$ mbar, respectively. The overall energy and angle resolutions are ~ 10 meV and 0.2°, respectively. Fresh ( 001 ) surfaces were obtained for measurements by *in situ* cleaving the crystals at low temperatures. No signatures of the surface aging were observed during measurements.

High-pressure electrical transport measurements were performed in a nonmagnetic diamond anvil cell ( DAC ) [40-43]. A cubic BN/epoxy mixture layer was inserted between BeCu gaskets and electrical leads. Four platinum sheet electrodes were touched to the sample for resistance measurements with the van der Pauw method. Pressure was determined by the ruby luminescence method [44]. An *in situ* high-pressure Raman spectroscopy investigation was performed using a Raman spectrometer ( Renishaw in-Via, UK ) with a laser excitation wavelength of 532 nm and a low-wavenumber filter. A symmetric DAC with anvil culet sizes of 200 μm was used, with silicon oil as pressure transmitting medium ( PTM ). *In situ* high-pressure XRD measurements were performed at beamline BL15U of Shanghai Synchrotron Radiation Facility ( x-ray wavelength $\lambda$ = 0.6199 Å ). A symmetric DAC with anvil culet sizes of 200 μm and Re gaskets were used. Silicon oil was used as the PTM and pressure was determined by the ruby luminescence method [44]. The two-dimensional diffraction images were analyzed using the FIT2D software [45]. Rietveld refinements of crystal structure under various pressures were performed using the GSAS and the graphical user interface EXPGUI [46, 47].

**RESULTS AND DISCUSISION**

Figure 1(a) shows the crystal structure of SrIn$_2$As$_2$, which adopts a hexagonal structure with a space group *P*6$_3$/*mmc*. The In and As atoms form two layers of a two-dimensional zigzag honeycomb lattice with mirror symmetry, forming [In$_2$As$_2$]$^{2-}$ polyanions through staggered stacking. Layers of Sr$^{2+}$ and [In$_2$As$_2$]$^{2-}$ are alternately stacked along the [0 0 1] direction. The single crystal XRD pattern of SrIn$_2$As$_2$ displays sharp [0 0 l] diffraction peaks [Figure 1(b)]. The extracted lattice parameter is $c$ = 18.08 Å, which is consistent with previous report [35]. Chemical composition analysis from EDX measurement on the single crystal reveals a good stoichiometry with atomic percentage of Sr : In : As = 19.96 : 40.93: 39.10, as shown in the inset of Figure 1(b). These characterizations indicate a high quality of our samples.

Next, we carried out transport measurements at ambient pressure. Figure 1(c) shows the temperature dependence of resistivity for SrIn$_2$As$_2$, which reveals metallic behavior with residual resistivity ratio ( RRR ) = 2.11. The longitudinal resistivity ( $\rho_{xx}$ ) and Hall resistivity ( $\rho_{yx}$ ) were measured with the magnetic field applied along *c*-axis at various temperature ranging from 1.8 K to 300 K, as shown in Figure S1. The positive magnetoresistance is observed in all the temperature range and reaches ∼ 0.47 % at 1.8 K and 9 T. At 1.8 K, $\rho_{yx}$ displays a linear dependence on the magnetic field with a positive slope, indicating that the hole carriers play a dominant role in the transport. This is also consistent with our ARPES measurements shown below. However, $\rho_{yx}$ shows almost temperature independent with a rather slight change over the measured temperature range. Hall resistivity is fitted using single-band model, i.e., $\rho_{yx} = R_H B$, where $R_H = 1/en$ is the Hall coefficient. Figure 1(d) shows the temperature dependence of carrier density *n* and carrier mobility *μ*. The carrier density *n* reaches 3.58×10$^{25}$ m$^{-3}$ and carrier mobility *μ* is estimated to be 343.94 cm$^2$·V$^{-1}$·s$^{-1}$ at 1.8 K, which are comparable with other topological materials [48].

Figure 2(a) and (b) display the measured constant-energy contours of SrIn$_2$As$_2$. The Fermi surface of SrIn$_2$As$_2$ is mainly composed of a circle encircling the $\bar{\Gamma}$ point. Inside the circle, some blurry features are also discernable. When going towards high-binding-energy regions, the circle keeps expanding, and circles from other adjacent Brillouin zones connect with each other to form a hexagon, consistent with the crystal symmetry of ( 001 ) surface. If cutting the Fermi surface along $\bar{\Gamma}$-$\bar{M}$ direction, one can see the sharp hole-like band dispersion marked as α that contributes to the circle Fermi surface, and multiple blurry bands residing between the two α bands [Figures 2(c) and (d)]. These dispersions exhibit qualitative agreement with previous theoretical calculations proposing SrIn$_2$As$_2$ is a dual topological insulator and hosts topological Dirac surface states on the ( 001 ) surface [Figure 2 (e)] [34]. However, we have not observed the whole Dirac surface states dispersion ( including the Dirac point ) as the crystals appear to be heavily *p*-doped compared to the calculations ( the red dotted line is the Fermi level from experiments in Figure 2(e) ). We tried *in-situ* potassium dosing but the shift-up of the Fermi level is quite limited. We ascribe the observed α bands to be the tail of the Dirac surface states. Their surface state nature is further verified by the detailed photon-energy dependence measurements as they show negligible k$_z$-dispersions [Figure 2(f)].

The Zintl compound SrIn$_2$As$_2$ with narrow band gap is extremely sensitive to external pressure. In addition, first principles calculations indicate that SrIn$_2$As$_2$ will undergo insulator–metal phase transition and topological quantum phase transition under pressure modulation. Indeed, the high-pressure approach has been widely employed in recent studies of topological materials and has led to many interesting results [49-51]. Hence, we are motived to investigate the effect of high pressure on the electrical transport properties of SrIn$_2$As$_2$. Figure 3 shows temperature dependence of the electrical resistivity *ρ*(T) of SrIn$_2$As$_2$ for pressure up to 70.9 GPa. At low pressure region, the *ρ*(T) first decrease with decrease temperature and reaches a minimum value. Then the *ρ*(T) gradually increases showing semiconducting-like behavior at 2.8 GPa with a negative d*ρ*/d*T* slope [Figure 3(a)]. The anomalous *ρ*(T) of SrIn$_2$As$_2$ implies that the carriers are weak localized at low temperature due to strong quantum coherence [38]. Upon further increasing the pressure, the resistivity at 300 K begins to drease rapidly and the semiconducting-like behavior is suppressed accompanied by a drop of *ρ*(T) at the lowest temperature ($T_{min}$ = 1.8 K), as show in the inset of Figure 3(a). With pressure increasing, the drop of $\rho$(T) becomes more pronounced and zero resistivity is achieved at low temperature for $P > 53.9$ GPa, indicating the emergence of superconducting transition. Moreover, the superconductivity with $T_c \sim 2.1$ K is robust and persists until the maximum pressure of 70.9 GPa [Figure 3(b)].

To gain insights into the superconducting transition, we applied external magnetic field on SrIn$_2$As$_2$ at 48.9 GPa and 59.8 GPa, respectively. It is clear that magnetic field could suppress superconducting transition ( Inset of Figure 3(d) ). $T_c$ gradually shifts toward lower temperature with the increase of the magnetic field. A magnetic field $\mu_0 H = 0.5$ T removes all signs of superconductivity above 1.8 K as shown in Figure 3(c). These results further confirm pressure-induced bulk superconductivity in SrIn$_2$As$_2$. The derived upper critical field $\mu_0 H_{c2}$ as a function of temperature can be well fitted using the empirical Ginzburg-Landau formula $\mu_0 H_{c2}(T) = \mu_0 H_{c2}(T)(1-t^2)/(1+t^2)$, where $t = T/T_c$ is the reduced temperature with zero-field superconducting $T_c$. The extrapolated upper critical field $\mu_0 H_{c2}(0)$ of SrIn$_2$As$_2$ from 90% $\rho_n$ criterion can reach 1.35 T at 59.8 GPa, which yields a Ginzburg-Landau coherence length $\xi_{GL}(0)$ of 15.60 nm. It is worth noting that the value of $\mu_0 H_{c2}(0)$ is well below the Pauli–Clogston limit.

To further identify the origin of superconductivity in SrIn$_2$As$_2$, *in situ* x-ray diffraction ( XRD ) measurements have been carried out to analysis the structure evolution under various pressures. Figure 4(a) shows the high-pressure synchrotron XRD patterns measured at room temperature up to 71.7 GPa. A representative refinement at 3.9 GPa is displayed in Figure S3(a). At lower pressure region, all the diffraction peaks can be indexed well to the hexagonal structure with a space group $P6_3/mmc$. The small peaks marked with an asterisk represent the signal of the rhenium gasket, which is caused by the dragging of the x-ray spot. Further increase of the applied pressure gives rise to the shift of the Bragg peaks to larger angles, due to the lattice shrinkage, as shown in the pressure dependences of the lattice parameters in Figure S3(b). The experience pressure-volume ($P$ - $V$) data can be well fitted using the third-order Birch-Murnaghan equation of state (EOS) [52], which obtained bulk modulus $K_0$ is 58.32 GPa with $V_0 = 277.91$ Å$^3$ and $K_0' = 5.80$. The structure of SrIn$_2$As$_2$ is robust and there is no structural phase transition until 23.2 GPa. However, when the pressure increases up to 30.8GPa, a broad peak appears around 14.4° in the diffraction patterns, indicating the application of high-pressure introduced amorphous phases in the SrIn$_2$As$_2$. A similar phenomenon was observed in some other materials [49, 51, 53-58]. It should be mentioned that the superconductivity is observed where pressure increase up to 30 GPa. To derive more structural information, high pressure *in situ* Raman spectroscopy measurements were carried out. As shown in Figure 4(c), Raman active modes are clearly observed at lower pressure regions. With increasing pressure, the profile of the spectra remains similar to that at lower pressure, whereas the observed modes exhibit blue shift, thus showing the normal pressure behavior. An abrupt disappearance of Raman peaks for pressure near to 23.7 GPa indicates the structural phase transition to an amorphous phase. The evolution of the Raman spectra is consistent with our synchrotron XRD patterns.

Several independent high-pressure transport measurements on SrIn$_2$As$_2$ single crystals provide consistent and reproducible results, confirming intrinsic superconductivity under high-pressure. From resistivity, XRD and Raman measurements, we can construct a *T-P* phase diagram of SrIn$_2$As$_2$ (Figure 5). In the lower pressure region, the $\rho$(T) decreases upon cooling in a metallic manner firstly and then undergoes a minimum value. Subsequently, the $\rho$(T) upturns rapidly just like a semiconductor characteristic. Such anomalous $\rho$(T) behavior due to weak localization effect provides a signal of quantum correction to conductance in SrIn$_2$As$_2$ [38]. Upon further increasing the pressure, the resistivity at room temperature as well as 3 K begins to drop rapidly and the semiconductor-like behavior is suppressed. Superconductivity is observed after the temperature dependence of $\rho$(T) changes from a semiconducting-like behavior to that of a metal. The $T_c$ of SrIn$_2$As$_2$ rises to 2.1 K and persists until the maximum pressure of this work. Above 30 GPa, *in situ* high pressure XRD demonstrated that a pressure-induced amorphization emerges. High pressure amorphous phase of SrIn$_2$As$_2$ retain after the pressure released, which confirmed by TEM measurements (Figures S3(c) and (d)). It is very interesting that an amorphous phase of SrIn$_2$As$_2$ could support superconductivity. This will be a stimulus for further research from an experimental and theoretical point of view.

**CONCLUSION**

In summary, we have successfully grown high-quality single crystal of the Zintl compound SrIn$_2$As$_2$. Combining the transports and ARPES measurements, we indicate that SrIn$_2$As$_2$ is a strong topological insulator. Moreover, we discovered pressure-induced superconductivity in SrIn$_2$As$_2$, and $T_c$ still persists until a maximum experimentally pressure. Our results demonstrate that the nontrivial topological state and the pressure-induced superconductivity were all observed in SrIn$_2$As$_2$. Thus, the Zintl $A$In$_2$$X_2$ ($A$ = Ca, Sr, and $X$ = P, As) family provides an excellent platform for further understanding the relationship between the topological phase and superconductivity.

**ACKNOWLEDGMENT**

This work was supported by the National Natural Science Foundation of China (Grant Nos. 52272265, U1932217, 11974246, 12004252), the National Key R&D Program of China (Grant No. 2018YFA0704300), and Shanghai Science and Technology Plan (Grant No. 21DZ2260400). H.F.Y. thanks the support from Shanghai Sailing Program (20YF1430500), the National Natural Science Foundation of China (Grant No. 12004248). Z.K.L. acknowledges the Technology Innovation Action Plan of the Science and Technology Commission of Shanghai Municipality with project number 20JC1416000. Z.W.W. thanks the support from the National Key R&D Program of China (Grant Nos. 2020YFA0308800 and 2022YFA1403400), the Natural Science Foundation of China (Grant No. 92065109), the Beijing Natural Science Foundation (Grant Nos. Z210006 and Z190006). The authors thank the Analytical Instrumentation Center (# SPST-AIC10112914), SPST, ShanghaiTech University and the Analysis and Testing Center at Beijing Institute of Technology for assistance in facility support. The authors thank the staffs from BL15U1 at Shanghai Synchrotron Radiation Facility for assistance during data collection.

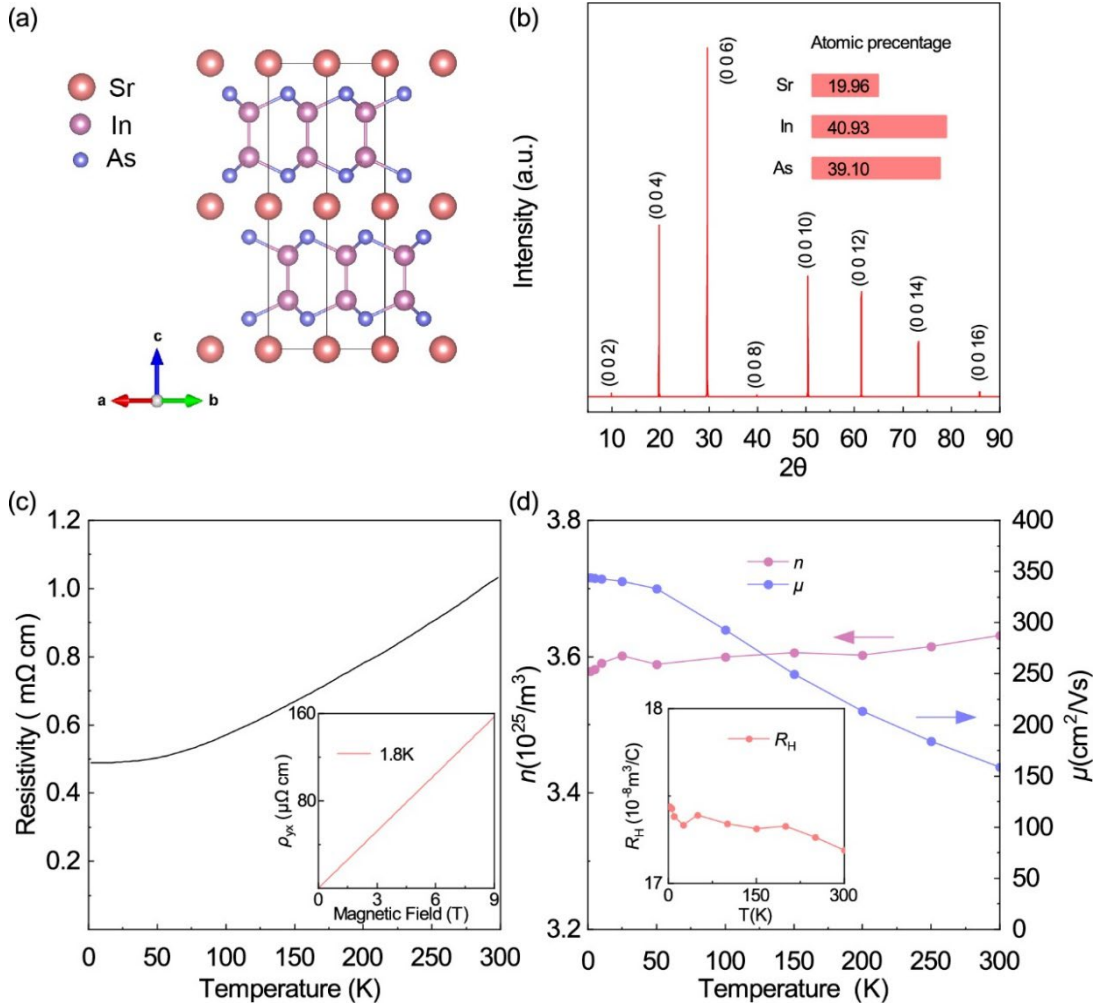

FIG. 1. (a) The crystal structure of SrIn$_2$As$_2$ with a space group $P6_3/mmc$. (b) The room-temperature x-ray diffraction peaks from the $ab$ plane of SrIn$_2$As$_2$ single crystal. Inset: the elemental content of SrIn$_2$As$_2$. (c) Resistivity of the SrIn$_2$As$_2$ single crystal from 1.8 K to 300 K. Inset: Hall resistivity [$\rho_{yx}(H)$] as a function of magnetic field at 1.8 K with magnetic field applied along the $c$-axis direction: (d) Carrier density $n$ and mobility $\mu$ of SrIn$_2$As$_2$ as a function of temperature, respectively. Inset: temperature dependence of Hall coefficient $R_H$ for SrIn$_2$As$_2$.

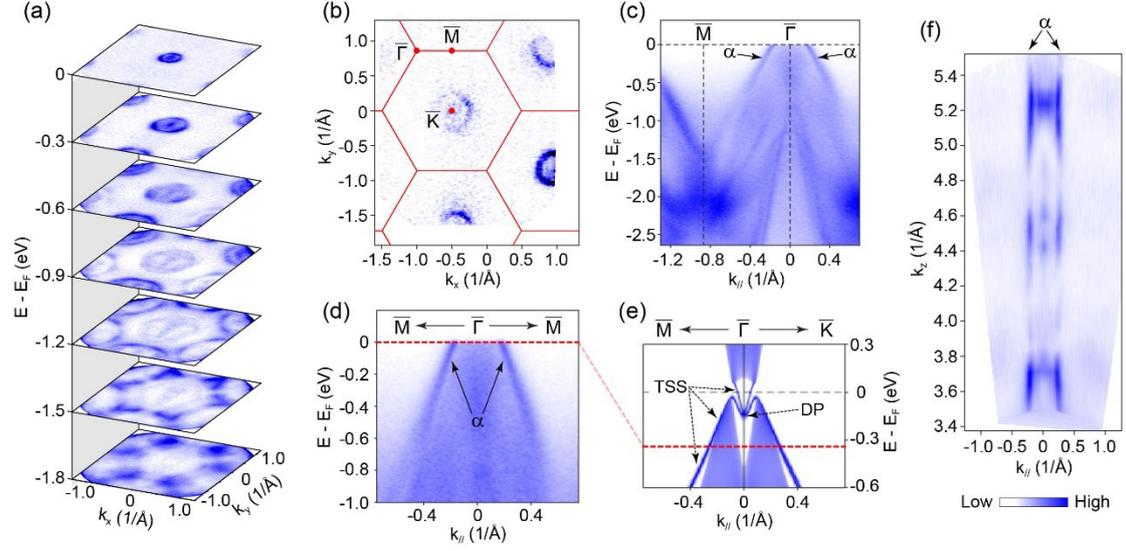

FIG. 2. ARPES measurements on SrIn$_2$As$_2$ crystals. (a) Stacking plot of constant energy contours measured in 98 eV photons with linearly horizontal polarization. (b) Fermi surface map across multiple Brillouin zones measured using 110 eV photons. The red lines represent the projected Brillouin zone. (c) High-symmetry band dispersion along the $\bar{\Gamma}$-$\bar{M}$ direction. (d) Zoom-in of the $\bar{\Gamma}$-$\bar{M}$ dispersion near the Fermi level. (e) Calculated $\bar{\Gamma}$-$\bar{M}$ dispersion clearly shows the Dirac surface states, reproduced from previous report [34]. TSS is short for topological surface states, DP is short for Dirac point. The dotted red line qualitatively marks the Fermi level from the ARPES measurements. (f) Photon energy dependent measurements apparently confirm the surface states nature of the α bands. The photon energy used ranges from 40 to 110 eV.

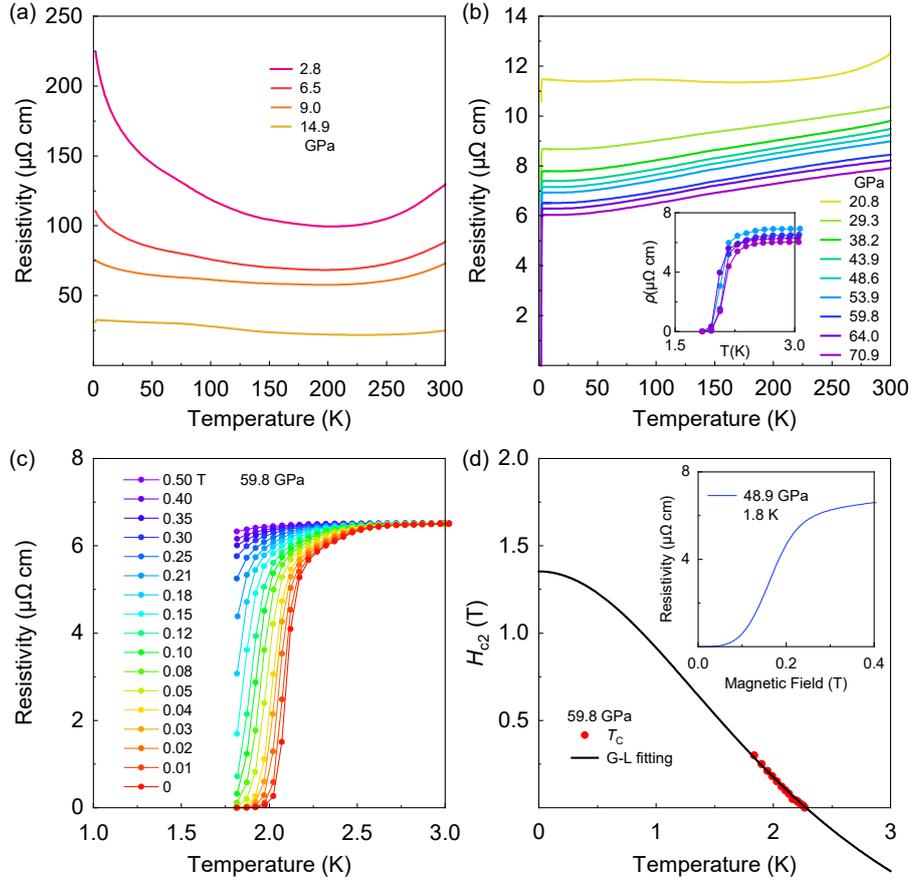

FIG. 3. Transport properties of SrIn$_2$As$_2$ as a function of pressure. (a) Electrical resistivity of SrIn$_2$As$_2$ as a function of temperature from 2.9 GPa to 14.9 GPa in run III. (b) Electrical resistivity of SrIn$_2$As$_2$ as a function of temperature from 20.8 GPa to 70.9 GPa in run III. Inset: Temperature-dependent resistivity of SrIn$_2$As$_2$ in the vicinity of the superconducting transition. (c) Resistivity as a function of temperature at the pressure of 59.8 GPa under different magnetic fields for SrIn$_2$As$_2$ in run III. (d) Temperature dependence of upper critical field for SrIn$_2$As$_2$ at 59.8 GPa. $T_c$ is determined as the 90% drop of the normal state resistivity. The solid lines represent the Ginzburg-Landau (G-L) fitting. The $\mu_0 H_{c2}(0)$ is 1.35 T. Inset: Resistivity as a function of magnetic field at 1.8 K. The superconductivity state is suppressed for field sufficiently away from 0.4 T.

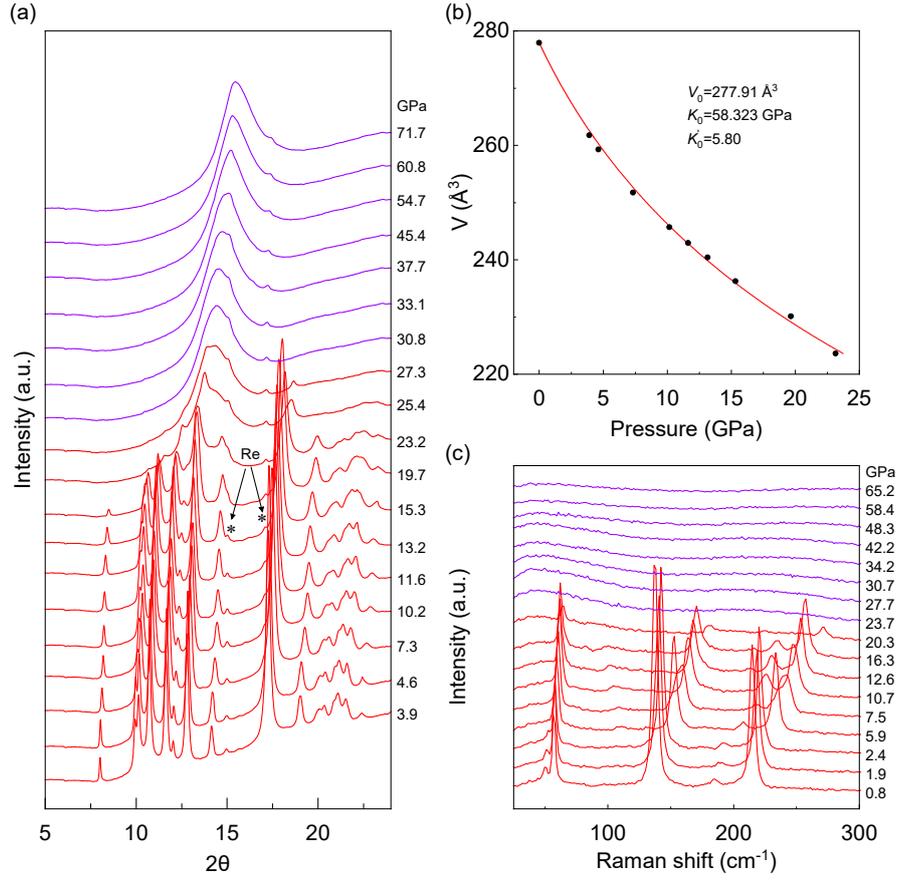

FIG. 4. (a) XRD patterns of SrIn$_2$As$_2$ under different pressures up to 71.7 GPa. (b) Variation in unit cell volume of SrIn$_2$As$_2$ with pressure at 300 K. Black circles: experiments. Solid lines are third-order Birch-Murnaghan fits to the data. The third-order Birch-Murnaghan equation of state (EOS) is given by $P = \frac{3}{2}K_0 \left[ \left(\frac{V_0}{V}\right)^{\frac{7}{3}} - \left(\frac{V_0}{V}\right)^{\frac{5}{3}} \right] \left\{ 1 + \frac{3}{4}(K_0' - 4)\left[ \left(\frac{V_0}{V}\right)^{\frac{2}{3}} - 1 \right] \right\}$. (c) Raman spectra of SrIn$_2$As$_2$ under pressure at room temperature.

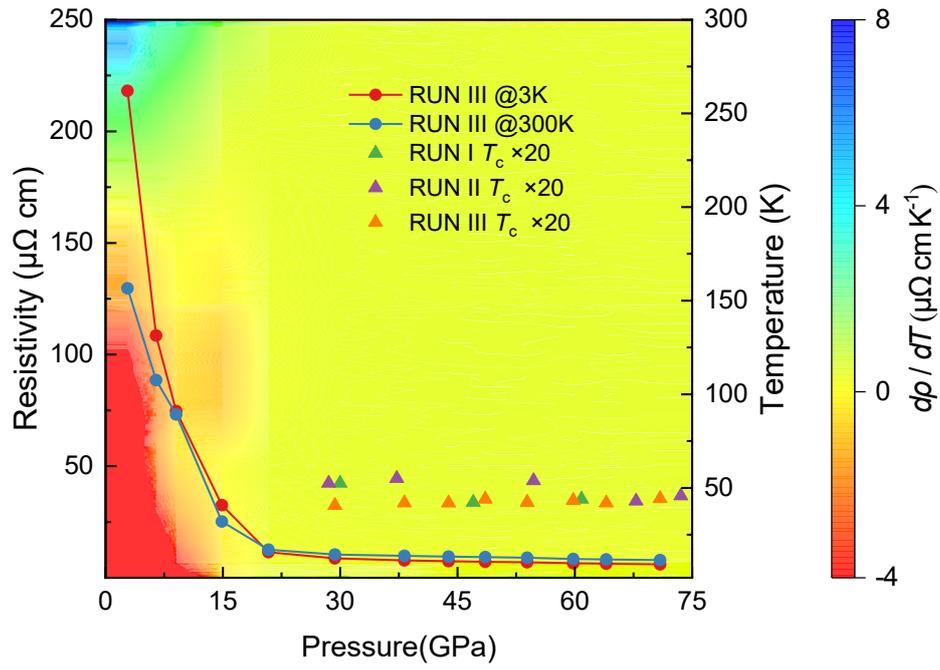

FIG. 5. Phase diagram of SrIn$_2$As$_2$. The red and blue dotted line plots represent the dependence of resistivity on temperature at 3 K and 300 K, respectively. Green, purple and orange triangle represent superconducting $T_c$ in different runs, respectively. The background represents the dependence of d$\rho$/d$T$ on temperature at different pressures